\documentclass[conference]{IEEEtran}
\IEEEoverridecommandlockouts
\usepackage{amsmath,amsfonts}
\usepackage{algorithmic}
\usepackage{algorithm}
\usepackage{array}
\usepackage[caption=false,font=normalsize,labelfont=sf,textfont=sf]{subfig}
\usepackage{textcomp}
\usepackage{stfloats}
\usepackage{url}
\usepackage{verbatim}
\usepackage{graphicx}
\usepackage{cite}
\usepackage{array}
\usepackage[caption=false,font=normalsize,labelfont=sf,textfont=sf]{subfig}
\usepackage{textcomp}
\usepackage{stfloats}
\usepackage{url}
\usepackage{verbatim}
\usepackage{graphicx}
\usepackage{xcolor}
\usepackage{cite}
\usepackage{tikz}
\usepackage{multirow}
\usepackage{pifont}
\newcommand{\cmark}{\ding{51}}
\newcommand{\xmark}{\ding{55}}
\def\BibTeX{{\rm B\kern-.05em{\sc i\kern-.025em b}\kern-.08em
    T\kern-.1667em\lower.7ex\hbox{E}\kern-.125emX}}
\begin{document}

\title{Decentralized PKI Framework for Data Integrity in Spatial Crowdsourcing Drone Services}

\author{
    \IEEEauthorblockN{Junaid Akram\IEEEauthorrefmark{1}, Ali Anaissi\IEEEauthorrefmark{1}\IEEEauthorrefmark{2}}
    \IEEEauthorblockA{\IEEEauthorrefmark{1}School of Computer Science, The University of Sydney, Camperdown NSW 2008, Australia\\
    \IEEEauthorrefmark{2}TD School, University of Technology Sydney, Ultimo NSW 2007, Australia\\
    Email: jakr7229@uni.sydney.edu.au, ali.anaissi@sydney.edu.au}
}

\maketitle

\begin{abstract}

In the domain of spatial crowdsourcing drone services, which includes tasks like delivery, surveillance, and data collection, secure communication is paramount. The Public Key Infrastructure (PKI) ensures this by providing a system for digital certificates that authenticate the identities of entities involved, securing data and command transmissions between drones and their operators. However, the centralized trust model of traditional PKI, dependent on Certificate Authorities (CAs), presents a vulnerability due to its single point of failure, risking security breaches. 
To counteract this, the paper presents D2XChain, a blockchain-based PKI framework designed for the Internet of Drone Things (IoDT). By decentralizing the CA infrastructure, D2XChain eliminates this single point of failure, thereby enhancing the security and reliability of drone communications. Fully compatible with the X.509 standard, it integrates seamlessly with existing PKI systems, supporting all key operations such as certificate registration, validation, verification, and revocation in a distributed manner. This innovative approach not only strengthens the defense of drone services against various security threats but also showcases its practical application through deployment on a private Ethereum testbed, representing a significant advancement in addressing the unique security challenges of drone-based services and ensuring their trustworthy operation in critical tasks.
\end{abstract}

\begin{IEEEkeywords}
Data Integrity, Decentralized Trust, Public Key Infrastructure, Spatial Crowdsourcing, Certificate Authority
\end{IEEEkeywords}

\section{Introduction}\label{sec:introduction}

In the contemporary digital era, the fusion of spatial crowdsourcing with the Internet of Drone Things (IoDT)\cite{akram2023chained, akram2022bc} marks a transformative approach to environmental monitoring\cite{munawar2022drone,munawar2022civil}, specifically in the context of Australian bushfire management\cite{munawar2022framework}. This innovative strategy employs drones and unmanned ground vehicles (UGVs), transcending traditional human-operated spatial crowdsourcing by offering expansive and real-time data collection in areas that are typically hazardous or unreachable.

Equipped with advanced communication technologies like WiFi/5G, drones and UGVs extend the capabilities of spatial crowdsourcing significantly. They excel in collecting vital sensory data from Points-of-Interest (PoIs), such as CCTV cameras and alarm sensors, covering larger areas with greater efficiency than human counterparts. To optimize data transmission and ensure high quality-of-service (QoS), the integration of the air-ground non-orthogonal multiple access (AG-NOMA) technique is proposed\cite{10492460, 10535995}. In this system, drones are pivotal in gathering uplink data and subsequently relaying it to UGVs. These UGVs serve a dual role: as mobile base stations for data decoding and as collectors of additional data from PoIs.

The application of IoDT, especially drones, in managing Australian bushfires, is not just about enhanced data collection\cite{akram2024DroneSSL, 10547221}. It represents the creation of a dynamic, real-time response system capable of adapting to the rapidly changing conditions of bushfires. Drones quickly detect and monitor bushfire developments, relaying critical information to UGVs and command centers, facilitating timely and effective decision-making and resource deployment.

Given the critical nature of the data involved in such emergency responses, ensuring its security and integrity is of utmost importance. Traditional digital security methods, centered around protocols like Transport Layer Security (TLS)\cite{2} and PKI\cite{3}, have exhibited certain vulnerabilities\cite{4,5,6}, as evidenced by breaches in entities like DigiNotar, StartCom, Comodo, and GoDaddy \cite{7,8,9}, exposing risks in centralized PKI. To address these concerns, we introduce D2XChain, a novel decentralized PKI framework specifically designed for IoDT. Incorporating blockchain technology, D2XChain distributes trust among Certificate Authorities (CAs), thereby reducing the single-point failure risks inherent in traditional PKI systems and enhancing the security of data transmission within the IoDT network\cite{16,17,18}.

D2XChain revolutionizes certificate management in the IoDT sphere, particularly for drone operations. It employs a unique consensus protocol, Proof-of-Service (PoSv), crafted for CA operations within IoDT networks. This approach not only fortifies the transmission and storage of data but also comprehensively manages all X.509 certificate tasks—registration, validation, revocation, and verification\cite{10,11,12,13,14,15}. By aligning with existing PKI standards, D2XChain ensures adaptability and robust security for diverse IoDT applications, including the critical function of monitoring and managing bushfires in Australia.

This integration of spatial crowdsourcing, IoDT, and secure data frameworks like D2XChain exemplifies how advanced technology can be harnessed to confront and manage environmental challenges. It highlights the potential of modern technological solutions in protecting natural landscapes and effectively responding to the increasing frequency and severity of bushfires in Australia.

In summary, our contributions are as follows:

\begin{itemize}
    \item We introduce D2XChain, a resilient blockchain-based PKI framework that decentralizes trust management for the IoDT, supported by a network of CA nodes.
    \item We introduce PoSv, an innovative consensus mechanism that D2XChain utilizes to facilitate all fundamental X.509 PKI operations within the IoDT sphere.
    \item D2XChain guarantees full adherence to the X.509 standard, managing certificates in their native format to ensure seamless integration with extant PKI systems.
    \item We conduct an exhaustive security analysis of D2XChain, evaluating it against the CIA triad—confidentiality, integrity, and availability—and perform a detailed threat analysis to mitigate PKI-related security challenges. The cryptographic rigor of our Drone Operator validation process is formally verified using the Verifpal tool.
    \item A meticulous security audit of the D2XChain smart contracts is carried out to affirm the framework's defenses against potential vulnerabilities.
    \item The performance of D2XChain is thoroughly assessed, with a focus on scalability, latency, throughput, and operational costs. We deploy D2XChain on a private Ethereum testbed, simulating various IoDT service scenarios to demonstrate its practicality and effectiveness.
\end{itemize}

This paper outlines the D2XChain framework, starting with PKI fundamentals and security issues in Section \ref{sec_2}, introducing D2XChain's design and architecture in Section \ref{sec_5}, detailing its operations in Section \ref{sec_6}, evaluating resilience and performance in Section \ref{sec_8}, and concluding with reflections and future directions in Section \ref{sec_9}.


\section{Related Work} \label{sec_2}

In the evolving landscape of PKI, researchers and developers are continually seeking solutions to enhance the system's security, trust, and efficiency. The focus has significantly shifted towards decentralizing the authority of CAs, incorporating logging mechanisms for transparent certificate verification, and leveraging blockchain technology for its immutability and distributed nature.

In the realm of CA-based trust models, innovations like the Web of Trust (WOT) have decentralized the trust by allowing network entities to authenticate each other, reducing the reliance on centralized CAs. ARPKI \cite{basin2014arpki} extends this concept by necessitating the joint action of multiple CAs for certificate signing, thus heightening security at the expense of efficiency and increasing vulnerability to Denial of Service (DoS) attacks. Toorani et al. \cite{toorani2021decentralized} introduced a blockchain-based PKI model, DBPKI, which eliminates traditional CA structures in favor of a decentralized approach using PKI nodes for key registration and revocation, supported by the Practical Byzantine Fault Tolerance (PBFT) protocol for consensus. Similarly, SCPKI \cite{al2017scpki} combines WOT with smart contracts, allowing entities to authenticate or endorse identity attributes via Ethereum addresses, bypassing traditional CA issuance processes.

Log-based PKI systems, exemplified by Google's Certificate Transparency (CT) \cite{laurie2014certificate}, employ immutable, append-only public ledgers to record certificate operations, ensuring validity through public log verification. This model aims for openness in CA certificate issuance, with systems like ARPKI, AKI \cite{kim2013accountable}, PoliCert \cite{szalachowski2014policert}, and CIRT \cite{ryan2013enhanced} building upon CT’s framework to enhance system integrity and traceability. However, these systems face challenges such as the centralization of log storage and the susceptibility to split-world attacks, where attackers present falsified versions of the log.

Blockchain-based PKI systems have emerged as a promising alternative, leveraging the decentralization and immutability of blockchain to address traditional PKI's limitations. Systems like CertCoin \cite{fromknecht2014decentralized} and BlockStack \cite{ali2016blockstack} base their PKI on Namecoin, linking users’ public keys to their identities on the blockchain. However, these solutions primarily focus on DNS architecture, leaving broader PKI aspects less addressed. PB-PKI \cite{axon2016pb} adds privacy considerations, crucial for applications in the Internet of Things (IoT) and the Internet of Vehicles (IoV). Cecoin \cite{qin2020cecoin}, a distributed certificate scheme inspired by Bitcoin, and CertChain \cite{chen2018certchain}, a publicly auditable blockchain-based TLS model, represent further advancements, though they also encounter challenges like managing certificate revocation and privacy concerns.

To address the scalability and efficiency challenges in blockchain-based PKI, researchers are exploring more efficient storage solutions and scalable blockchain architectures. The integration of off-chain storage mechanisms for certificate data and advancements in consensus mechanisms are among the solutions considered to reduce blockchain size and enhance performance.

The InterPlanetary File System (IPFS) has also been explored for certificate storage in PKI systems, offering a decentralized approach to improve privacy and cross-domain authentication. Systems like XAuth \cite{chen2021xauth} and SCPKI use IPFS to store attribute data securely, while blockchain-based identity management and authentication architectures for VANETs utilize IPFS for direct certificate retrieval. However, these IPFS-based systems still face challenges in protecting the privacy of issuing CAs and preventing the distribution of certificates by dishonest CAs.

As the PKI landscape continues to evolve, the integration of blockchain technology, decentralized storage solutions like IPFS, and innovative trust models promise to address the existing challenges while introducing new opportunities for secure and efficient digital certificate management.

Notwithstanding these advancements, the current IPFS certificate storage systems fall short in protecting the issuing CAs' privacy. This error exposes them to focused attacks and does not stop dishonest CAs from distributing certificates in the network incorrectly.

\section{D2XChain: A Blockchain-based PKI Framework for IoDT}\label{sec_5}

                      This section outlines D2XChain's design objectives, system and adversary models, and operational components.

\subsection{Design Goals}\label{sec_5.1}

D2XChain aims to create a robust, decentralized PKI framework for IoDT, compliant with the X.509 standard. Its design goals are:

\begin{itemize}
    \item \textbf{X.509 Compatibility:} Ensuring all key PKI functions for IoDT, leveraging blockchain for certificate management.
    \item \textbf{Decentralized Trust:} Distributing trust to avoid single points of failure and enhancing overall security.
    \item \textbf{MITM Attack Mitigation:} Implementing stringent security protocols during certificate validation.
    \item \textbf{Automation:} Utilizing smart contracts and consensus protocols for automated certificate lifecycle management.
    \item \textbf{Transparency:} Maintaining a transparent and immutable record of certificate histories.
\end{itemize}

\subsection{System Model}\label{sec_5.2}

D2XChain integrates blockchain's architectural and consensus features for IoDT PKI, comprising four main operations:

\begin{itemize}
    \item \textbf{Registration:} Requesting new certificates for IoDT network entities.
    \item \textbf{Validation:} Verifying certificate requests to ensure legitimacy and origin.
    \item \textbf{Revocation:} Invalidating certificates upon security compromise or end of service.
    \item \textbf{Verification:} Ensuring certificates are valid, unexpired, and unrevoked.
\end{itemize}

\subsubsection{D2XChain Entities}\label{sec_5.2.1}

D2XChain includes three main entities, each integral to the IoDT ecosystem:

\begin{itemize}
    \item \textbf{Drone Operators:} Registered members managing drones and associated TLS certificates, with unique key pairs.
    \item \textbf{Service Validators:} Nodes responsible for maintaining the blockchain ledger state and validating service requests.
    \item \textbf{IoDT Users:} Individuals interacting with drone services, involved in verifying communication security.
\end{itemize}

\subsubsection{D2XChain Transactions}\label{sec_5.2.3}

D2XChain enables operations like certificate requests (CRTs) through transactions, including initial registration and revocation requests. Transactions comprise a header, indicating the transaction type (\textit{CRT initial} or \textit{CRT revoke}), and a body containing the drone's details. Validated transactions are added as blocks to the D2XChain, each block holding a single transaction for clear drone certificate status recording.

\subsubsection{D2XChain Components}

Key components of D2XChain in the IoDT ecosystem are:

\begin{itemize}
    \item \textbf{Pending Transactions Pool:} Holds unvalidated transactions from drone operators.
    \item \textbf{D2XChain Ledger:} Stores validated transactions as blocks in chronological order, maintaining two states: global (all drone certificates) and individual (each drone).
    \item \textbf{D2XChain Client Plugin:} Integrated into the drone operator's interface for certificate verification, ensuring real-time ledger connectivity.
\end{itemize}

\subsubsection{Adversary Model}

The adversary model for D2XChain considers potential threats like malicious service validators and MITM attacks. It relies on a sufficient number of trustworthy validators adhering to the consensus protocol, ensuring system integrity despite potential risks from various participants, including drone operators and end-users.

\section{D2XChain Architecture}\label{sec_6}

This section describes D2XChain's operational mechanics, covering the interplay between its operations, components, entities, and transactions, and the process of transforming transactions into blocks within the IoDT context.

\subsection{D2XChain Workflow}\label{sec_6.1}

\begin{figure}
    \centering
    \includegraphics[width=\columnwidth]{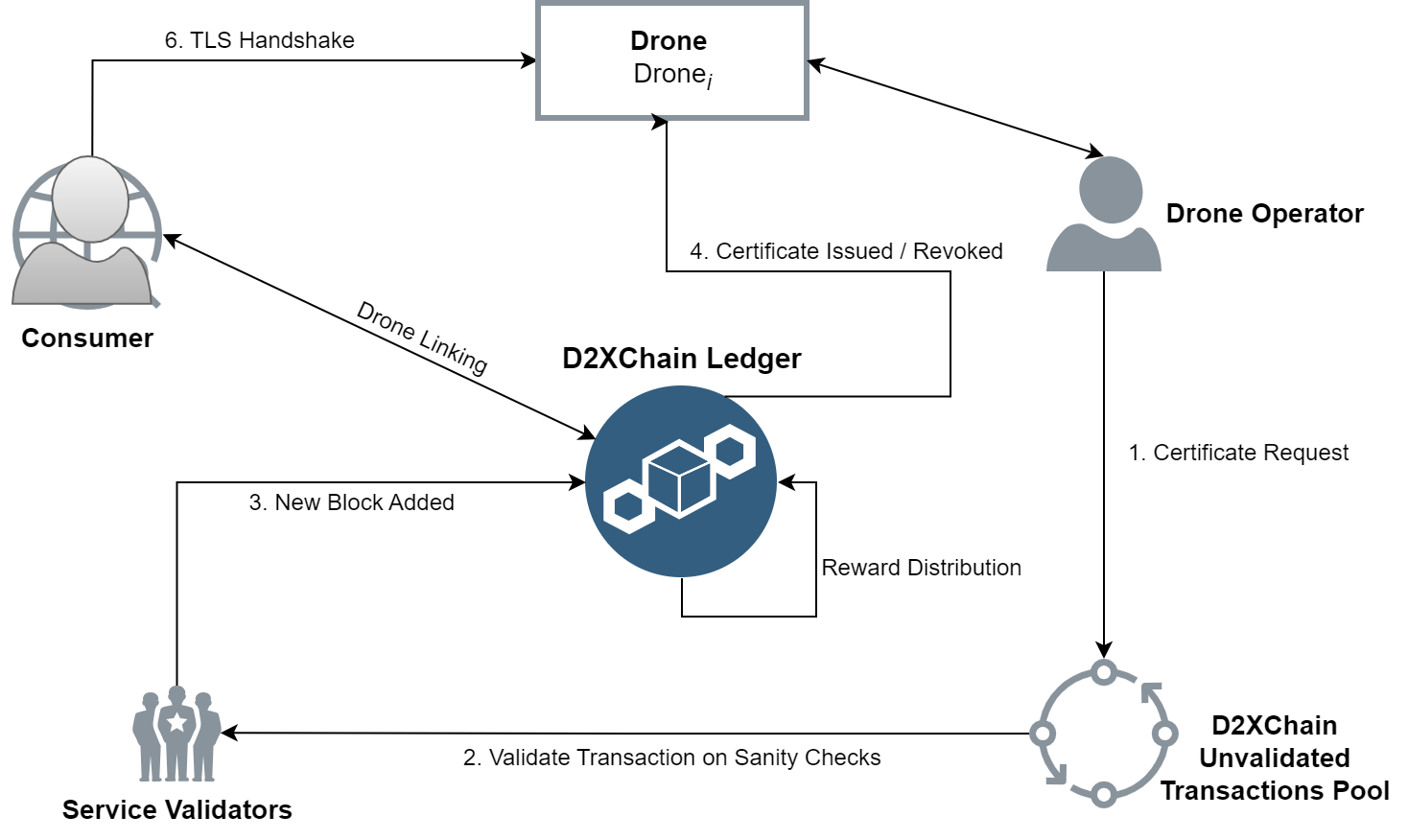}
    \caption{D2XChain Workflow: a structured approach for D2XChain components, entities, and operations to communicate with one other inside the IoDT.}
    \label{fig_4}
\end{figure}

As illustrated in Figure \ref{fig_4}, D2XChain's workflow involves a drone operator initiating a certificate request (CRT), processed by service validators and placed in the pending transactions pool. Validators authenticate transactions and add them to the D2XChain ledger as blocks upon majority approval. Transaction types include CRT initial (certificate issuance) and CRT revoke (certificate annulment). Validators receive rewards for their contribution, while the D2XChain client plugin verifies certificate validity during initial communications.

\subsection{Blocks in D2XChain Ledger}\label{sec_6.2}

A ratified transaction becomes a block in the D2XChain ledger upon consensus among service validators. This subsection explores the block structure and the D2XChain ledger's design principles, emphasizing transparency, immutability, and traceability. Modifications in drone services are swiftly shared and traceable. Each D2XChain block consists of three parts: header, body, and a footer with validator signatures, ensuring secure and transparent record-keeping of drone service changes.

D2XChain blocks are composed of three sections: Header, Body, and Footer, each serving specific functions.

\begin{enumerate}
    \item \textbf{Header:} Integrates unique block identifiers in the D2XChain ledger, with fields for Serial Number, CRT Type, Global State (linking all drone services via hash pointers), and Service-Specific State (maintaining a chronological chain for each drone service).

    \item \textbf{Body:} Houses service-related data, including Drone Name, Public Key of the Drone Operator, Signature of the Drone Operator, and the Expiry Date of the certificate.

    \item \textbf{Footer:} Contains the Multi-Signature of service validators, signifying their consensus on the block's validity.
\end{enumerate}

The multi-signature block, denoted as $\psi_i$, is authenticated by $n-1$ service validators, as expressed in Equation \ref{eq_2}.

\begin{equation}\label{eq_2}
    \psi_i \equiv Tx(C_1(sig_1) + C_2(sig_2) + \ldots + C_{n-1}(sig_{n-1}))
\end{equation}

\begin{figure}
    \centering
    \includegraphics[width=\columnwidth]{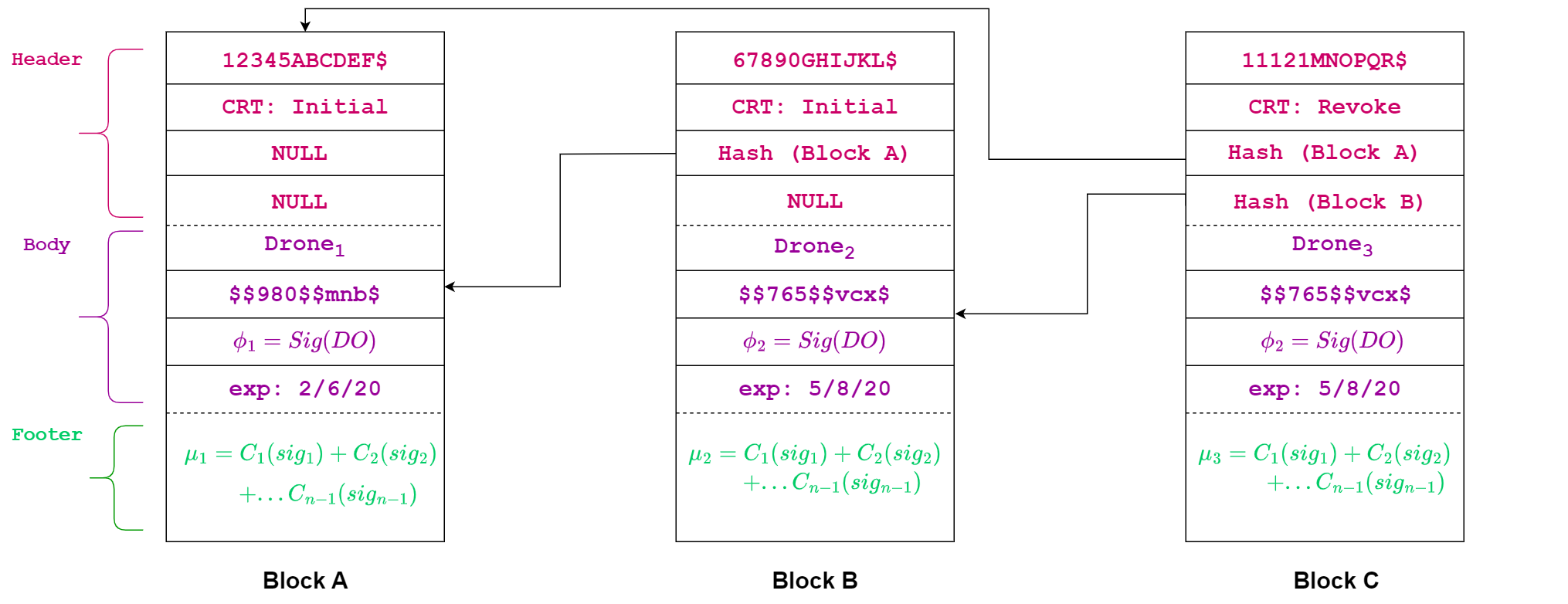}
    \caption{Organization of blocks in the D2XChain ledger for IoDT services.}
    \label{fig_5}
\end{figure}

In addition, Figure \ref{fig_5} shows how blocks are arranged inside the D2XChain ledger.

\section{D2XChain and PKI Functionalities}\label{sec_7}

The four main functions of D2XChain—registration, validation, revocation, and verification—are examined in this section. They are all framed under the PoSv model. PoSv is essential for upholding the integrity of PKI functioning and verifying the choice of service validators, as well as providing a means of compensation for their sincere involvement.

\subsection{Registration and Validation of Service Certificates}\label{sec_7.1}

A Drone Operator initiates a certificate request via a CRT initial transaction for their IoDT service. The service validators perform independent sanity checks to authenticate these transactions, ensuring the integrity and legitimacy of the PKI functionalities within D2XChain.

    \begin{algorithm}[t]
    \caption{Registration and Validation of Service Certificate}\label{alg:alg1}
    \begin{algorithmic}
    \STATE 
    \STATE {\textbf{Input}: CSR Hash, drone Name (DN), UserID}
    \STATE {\textbf{Output}: Certificate issued to Drone Operator and transaction is added in D2XChain Ledger}
    \STATE {\textbf{Procedure} Registration and Validation}
    \STATE \hspace{0.5cm}$T_x$ = CreateTx(DN,CSR)
    \STATE \hspace{0.5cm}$h_1\leftarrow$ hash ($SV_{\text{pk}}+Tx$)
    \STATE \hspace{0.5cm}$T\leftarrow \text{Sign}_{\text{SK}}^{\text{SV}}(h_1)$ 
    \STATE \hspace{0.5cm}$\text{DO}\leftarrow \text{T}$
    \STATE \hspace{0.5cm}$T_0\leftarrow \text{Sign}_{\text{SK}}^{\text{DO}}(T)$
    \STATE \hspace{0.5cm}PlaceToken $(T_0)$
    \STATE \hspace{0.5cm}$T_1,status\leftarrow$ RetrieveToken()
    \STATE \hspace{0.5cm}\textbf{if} status == 1 \textbf{then}
    \STATE \hspace{1cm}Success, continue
    \STATE \hspace{0.5cm}$T_2\leftarrow \text{Decrypt}_{\text{PK}}^{\text{DO}}(T_1)$
    \STATE \hspace{0.5cm}\textbf{if}  $T_2$ equals $T$ \textbf{then}
    \STATE \hspace{1cm}Success, continue
    \STATE \hspace{0.5cm}$\text{Sign}_{\text{Tx}}\leftarrow \text{Sign}_{\text{SK}}^{\text{SV}}(T_x).$
    \STATE \hspace{0.5cm}Propose ($\text{Sign}_{\text{Tx}}$)
    \end{algorithmic}
    \end{algorithm}

D2XChain employs a series of steps to validate certificate requests by Drone Operators:

\begin{enumerate}
    \item \textbf{Key Verification}: Confirms the match between the Drone Operator's public key and signature, verifying private key possession.
    \item \textbf{Domain Existence Check}: Ensures the drone's public key is not already present in the D2XChain ledger to avoid impersonation.
    \item \textbf{Drone Operator Verification}: Requires a signed token `$t_0$` placed on the DNS to authenticate the Drone Operator's legitimacy.

    \begin{equation}\label{eq_3}
        t_0\equiv \text{Sign}_{\text{SK}}^{\text{SV}}\left(\text{hash}\left(\text{Tx}+SV_{\text{pk}}\right)\right)
    \end{equation}

    \item \textbf{Token Placement and Retrieval}: Involves the Drone Operator signing and placing the token, and the service validator retrieving and validating it.

    \begin{equation}\label{eq_4}
        \phi_0\equiv \text{Sign}_{\text{SK}}^{\text{DO}}(t_0)
    \end{equation}

\end{enumerate}

\begin{figure}
    \centering
    \includegraphics[width=0.7\columnwidth]{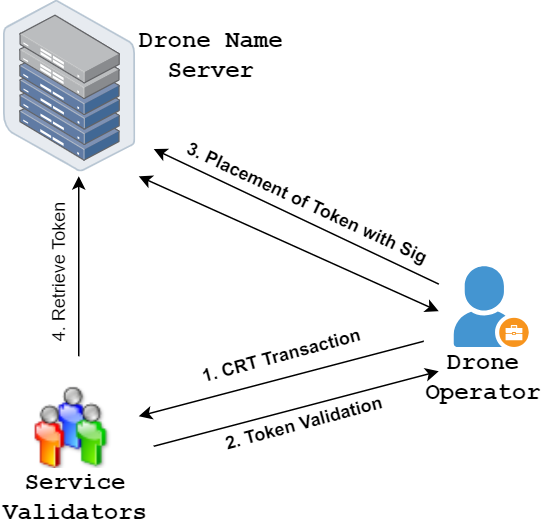}
    \caption{Drone Operator validation process in D2XChain.}
    \label{fig_6}
    \end{figure}

Successful completion of these checks leads to the validation of the transaction by the service validator. The detailed process is outlined in Algorithm \ref{alg:alg1} and Figure \ref{fig_6}.

D2XChain's revocation process involves several steps to ensure the authenticity and integrity of the revocation request:

\begin{enumerate}
    \item \textbf{Key Verification}: Checks the alignment of the public key and signature to affirm the Drone Operator's control over the corresponding private key.
    \item \textbf{Drone Existence Check}: Confirms the presence of the drone-public key binding in the ledger, which is essential for a revocation request.
    \item \textbf{Revoke Request Validation}: The Drone Operator generates a token combining the hash of the last transaction for the drone, the current request, and their public key.

    \begin{equation}\label{eq_5}
        t_1\equiv \text{hash}\left(Tx_{\text{prev}}^d+Tx_{\text{cur}}^d+DO_{\text{pk}}\right)
    \end{equation}

    \item \textbf{Token Placement and Validation}: The Drone Operator signs and places the token on the IoDT platform, after which the service validator retrieves and validates it.

    \begin{equation}\label{eq_6}
        \phi_1\equiv \text{Sign}_{\text{SK}}^{\text{DO}}(t_1)
    \end{equation}

    \item Upon successful completion of these steps, the service validator officially endorses the revocation transaction.
\end{enumerate}

The detailed procedure for token generation, placement, and validation is elaborated in Algorithm \ref{alg:alg2}.

\begin{algorithm}[t]
\caption{Revocation of Service Certificate}\label{alg:alg2}
\begin{algorithmic}
\STATE 
\STATE {\textbf{Input}: Previous Transaction Hash, Current Transaction, drone Name (DN), UserID}
\STATE {\textbf{Output}: Certificate revocation and transaction update in D2XChain Ledger}
\STATE {\textbf{Procedure} Revocation of Certificate}
\STATE \hspace{0.5cm}$T_x$ = CreateTx(DN,UserID)
\STATE \hspace{0.5cm}$h_2\leftarrow$ hash ($Tx_{\text{prev}}^d+Tx_{\text{cur}}^d+DO_{\text{pk}}$)
\STATE \hspace{0.5cm}$T\leftarrow \text{Sign}_{\text{SK}}^{\text{DO}}(h_2)$ 
\STATE \hspace{0.5cm}PlaceToken $(T)$
\STATE \hspace{0.5cm}$T_1,status\leftarrow$ RetrieveToken()
\STATE \hspace{0.5cm}\textbf{if} status == 1 \textbf{then}
\STATE \hspace{1cm}Success, continue
\STATE \hspace{0.5cm}$h_3\leftarrow$ hash ($Tx_{\text{prev}}^d+Tx_{\text{cur}}^d+DO_{\text{pk}}$)
\STATE \hspace{0.5cm}\textbf{if}  $h_3$ equals $T_1$ \textbf{then}
\STATE \hspace{1cm}Success, continue
\STATE \hspace{0.5cm}$\text{Sign}_{\text{Tx}}\leftarrow \text{Sign}_{\text{SK}}^{\text{SV}}(T_x).$
\STATE \hspace{0.5cm}Propose ($\text{Sign}_{\text{Tx}}$)
\end{algorithmic}
\end{algorithm}

All Service Validators in the IoDT framework follow the same protocol to validate revocation requests. A consensus of more than 50\% of SVs leads to the appending of a CRT revoke block to the D2XChain ledger, invalidating the certificate within IoDT. These protocols operate within a specialized Ethereum network sandbox tailored for IoDT services.

\subsection{Verification of Certificate}\label{sec_7.4.2}

\begin{figure}
    \centering
    \includegraphics[width=0.8\columnwidth]{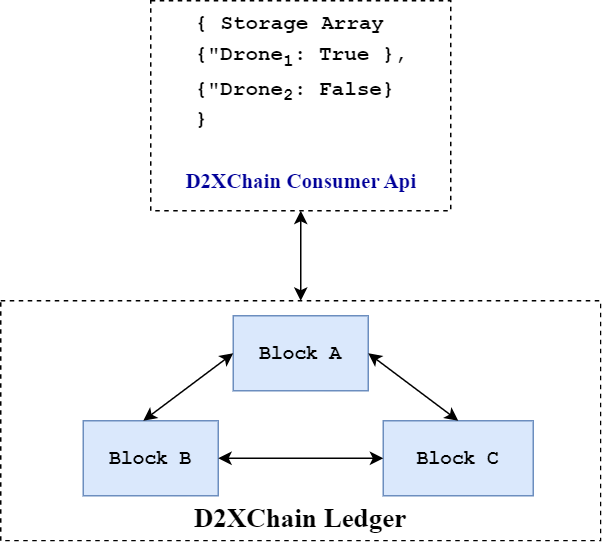}
    \caption{Certificate verification via the D2XChain client plugin in IoDT.}
    \label{fig_7}
\end{figure}

Certificate verification in D2XChain is managed by a client plugin within the IoDT ecosystem. It checks the validity of certificates against a synchronized storage array in the plugin, aligning with the D2XChain ledger:

\begin{itemize}
    \item A value of true indicates an active, valid certificate, as seen with "$Drone_1$" in Fig. \ref{fig_7}.
    \item A value of false signifies an unregistered, revoked, or expired certificate, exemplified by "$Drone_2$" in Fig. \ref{fig_7}.
\end{itemize}

This verification mechanism is pivotal for ensuring trusted and secure communication within the IoDT, allowing for the quick identification and invalidation of compromised or outdated certificates.

\subsection{Service Validator Selection and Incentivization}\label{sec_7.4}

To enhance integrity in the IoDT framework, an incentive mechanism is introduced for Service Validators. Drone Operators (DOs) pay fees for certificate operations, which are then distributed to SVs as a reward for their validation efforts, promoting system integrity. The specifics of fee allocation are outside this paper's scope.

D2XChain addresses the risk of centralized trust in conventional PKI by employing a decentralized consortium of SVs. The system's trust relies on the majority of SVs being honest, reducing the risk associated with single CA compromises. The framework suggests enlisting established root CAs as SVs and adjusting their number for optimal scalability and performance. Details on SV registration and miner certificate issuance are reserved for future studies.

\section{Evaluation and Discussion}\label{sec_8}

We now assess the D2XChain framework against its security objectives and adversarial model, alongside a performance analysis within the IoDT context.

\subsection{Security Assessment}\label{sec_8.1}

D2XChain's security is evaluated against the CIA triad model \cite{47}, focusing on confidentiality, integrity, and availability in the IoDT ecosystem. Detailed discussions of these security aspects will follow in subsequent subsections.

D2XChain's security within the IoDT framework focuses on confidentiality, integrity, and availability:

\begin{itemize}
    \item \textbf{Confidentiality}: Ensured through asymmetric cryptography between SVs and DOs, with the safeguarding of private keys off-chain and transaction validation without knowledge of peers' actions.
    \item \textbf{Integrity}: Maintained in SV-DO communication via signed tokens in DNS records and accountable SVs under the consensus mechanism, compatible with both secure and traditional DNS implementations.
    \item \textbf{Availability}: High availability achieved through a distributed ledger resistant to DoS attacks, bolstered by the geographical dispersion and redundancy of SV nodes.
\end{itemize}

\subsection{Formal Modeling and Verification for IoDT}\label{sec_8.3}

\begin{figure}
    \centering
\tikzset{every picture/.style={line width=0.75pt}}
\begin{tikzpicture}[x=0.55pt,y=0.6pt,yscale=-1,xscale=1]
\draw    (388.67,38.33) -- (388.67,624.52) ; 
\draw    (224.67,38.33) -- (224.67,624.52) ; 
\draw  [fill={rgb, 255:red, 0; green, 0; blue, 0 }  ,fill opacity=1 ] (178.67,8.67) -- (270.33,8.67) -- (270.33,37.52) -- (178.67,37.52) -- cycle ; 
\draw  [fill={rgb, 255:red, 0; green, 0; blue, 0 }  ,fill opacity=1 ] (355,8.67) -- (422,8.67) -- (422,37.52) -- (355,37.52) -- cycle ;
\draw  [fill={rgb, 255:red, 230; green, 221; blue, 221 }  ,fill opacity=1 ] (172.33,51.67) -- (276.67,51.67) -- (276.67,86.85) -- (172.33,86.85) -- cycle ; 
\draw  [fill={rgb, 255:red, 230; green, 221; blue, 221 }  ,fill opacity=1 ] (336,103) -- (440.33,103) -- (440.33,138.19) -- (336,138.19) -- cycle ; 
\draw  [fill={rgb, 255:red, 230; green, 221; blue, 221 }  ,fill opacity=1 ] (112,209.85) -- (335.67,209.85) -- (335.67,288.85) -- (112,288.85) -- cycle ; 
\draw  [fill={rgb, 255:red, 230; green, 221; blue, 221 }  ,fill opacity=1 ] (256.67,331.52) -- (519.67,331.52) -- (519.67,424.85) -- (256.67,424.85) -- cycle ;
\draw  [fill={rgb, 255:red, 230; green, 221; blue, 221 }  ,fill opacity=1 ] (85.67,468.19) -- (362.33,468.19) -- (362.33,517.52) -- (85.67,517.52) -- cycle ;
\draw  [fill={rgb, 255:red, 230; green, 221; blue, 221 }  ,fill opacity=1 ] (231,561.52) -- (545.67,561.52) -- (545.67,610.85) -- (231,610.85) -- cycle ; 
\draw  [fill={rgb, 255:red, 0; green, 0; blue, 0 }  ,fill opacity=1 ] (178.67,625.33) -- (270.33,625.33) -- (270.33,654.19) -- (178.67,654.19) -- cycle ; 
\draw  [fill={rgb, 255:red, 0; green, 0; blue, 0 }  ,fill opacity=1 ] (355.67,624.52) -- (422.67,624.52) -- (422.67,653.38) -- (355.67,653.38) -- cycle ;
\draw    (224.67,166.67) -- (383,166.67) ;
\draw [shift={(386,166.67)}, rotate = 180] [fill={rgb, 255:red, 0; green, 0; blue, 0 }  ][line width=0.08]  [draw opacity=0] (6.25,-3) -- (0,0) -- (6.25,3) -- cycle    ; 
\draw    (224.67,316.67) -- (383,316.67) ;
\draw [shift={(386,316.67)}, rotate = 180] [fill={rgb, 255:red, 0; green, 0; blue, 0 }  ][line width=0.08]  [draw opacity=0] (6.25,-3) -- (0,0) -- (6.25,3) -- cycle    ;
\draw    (224.67,545.67) -- (383,545.67) ;
\draw [shift={(386,545.67)}, rotate = 180] [fill={rgb, 255:red, 0; green, 0; blue, 0 }  ][line width=0.08]  [draw opacity=0] (6.25,-3) -- (0,0) -- (6.25,3) -- cycle    ; 
\draw    (386,194.67) -- (227.67,194.67) ;
\draw [shift={(224.67,194.67)}, rotate = 360] [fill={rgb, 255:red, 0; green, 0; blue, 0 }  ][line width=0.08]  [draw opacity=0] (6.25,-3) -- (0,0) -- (6.25,3) -- cycle    ;
\draw    (386,452.67) -- (227.67,452.67) ;
\draw [shift={(224.67,452.67)}, rotate = 360] [fill={rgb, 255:red, 0; green, 0; blue, 0 }  ][line width=0.08]  [draw opacity=0] (6.25,-3) -- (0,0) -- (6.25,3) -- cycle    ;
\draw (225.5,24.09) node  [font=\scriptsize,color={rgb, 255:red, 255; green, 255; blue, 255 }  ,opacity=1 ] [align=left] {Drone\_Operator};
\draw (389.5,24.09) node  [font=\scriptsize,color={rgb, 255:red, 255; green, 255; blue, 255 }  ,opacity=1 ] [align=left] {Service\_Validator};
\draw (225.5,70.26) node  [font=\scriptsize] [align=left] {knows private a \\ga = G\textasciicircum a};
\draw (389.17,121.59) node  [font=\scriptsize] [align=left] {knows private b \\gb = G\textasciicircum b};
\draw (224.83,250.35) node  [font=\scriptsize] [align=left] {generates csr\\generates dronename\\certreq = CONCAT (csr, dronename)\\encreq=PKE\_ENC (gb, certreq)\\signature = SIGN(a, encreq)};
\draw (389.17,379.19) node  [font=\scriptsize] [align=left] {vrf = SIGNVERIF (ga, encreq, signature)?\\d = PKE\_DEC (b, encreq)\\generates tokentx\\token = CONCAT (gb, tokentx)\\enctoken = PKE\_ENC (ga, token)\\tokensig = SIGN (b, enctoken)};
\draw (225,493.19) node  [font=\scriptsize] [align=left] {vrfy = SIGNVERIF (gb, enctoken, tokensig)?\\enctokendo = PKE\_ENC (gb, enctoken)\\sigtokendo = SIGN (a, enctokendo)};
\draw (389.33,587.19) node  [font=\scriptsize] [align=left] {verify = SIGNVERIF (ga, enctokendo, sigtokendo)?\\initialtoken = PKE\_DEC (b, enctokendo)\\\_ = ASSERT (initialtoken, enctoken)};
\draw (225.5,640.76) node  [font=\scriptsize,color={rgb, 255:red, 255; green, 255; blue, 255 }  ,opacity=1 ] [align=left] {Drone\_Operator};
\draw (390.17,639.95) node  [font=\scriptsize,color={rgb, 255:red, 255; green, 255; blue, 255 }  ,opacity=1 ] [align=left] {Service\_Validator};
\draw (306.33,543.67) node [anchor=south] [inner sep=0.75pt]  [font=\small] [align=left] {enctokendo, sigtokendo};
\draw (306.33,450.67) node [anchor=south] [inner sep=0.75pt]  [font=\small] [align=left] {enctoken, tokensig};
\draw (306.33,314.67) node [anchor=south] [inner sep=0.75pt]  [font=\small] [align=left] {encreq, signature};
\draw (306.33,192.67) node [anchor=south] [inner sep=0.75pt]  [font=\small] [align=left] {[gb]};
\draw (306.33,164.67) node [anchor=south] [inner sep=0.75pt]  [font=\small] [align=left] {[ga]};
\end{tikzpicture}
    \caption{Verifpal model for certificate issuance.}
    \label{fig_8}
    \end{figure}
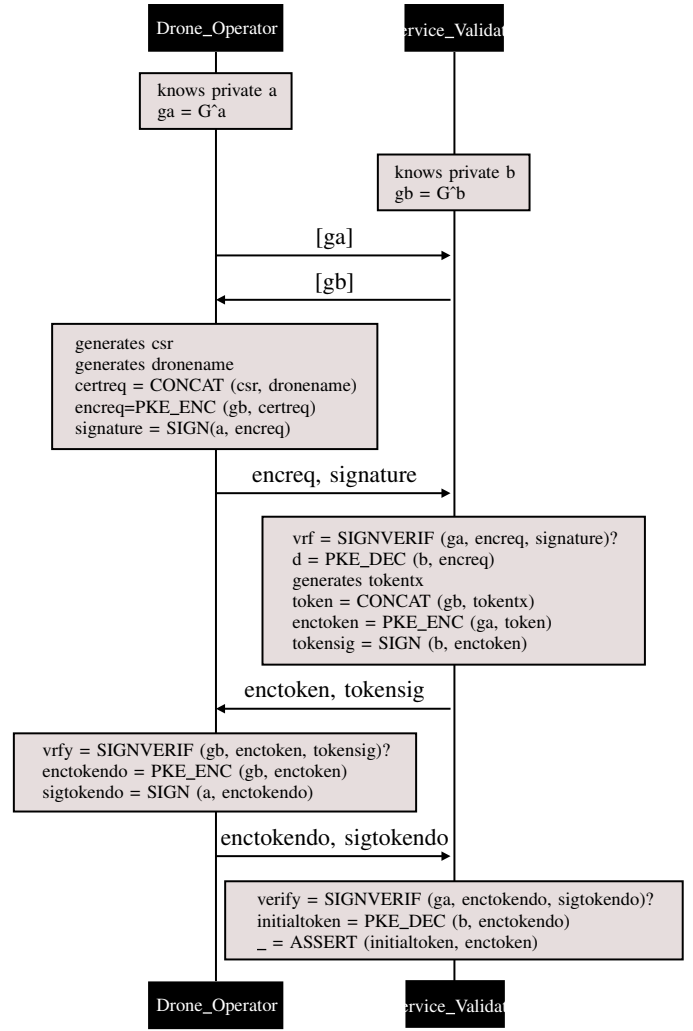

The cryptographic properties of D2XChain for IoDT were formally verified using Verifpal \cite{49}. The detailed Verifpal model for certificate issuance is given in Figure \ref{fig_8}. The tool's proficiency in analyzing complex protocols and assessing post-compromise scenarios is particularly effective in the IoDT context. Verifpal's symbolic formal verification approach ensures security in communication sequences between SVs and DOs, with a focus on confidentiality and authentication. The comprehensive model and verification process is depicted in Figure 8.

\subsubsection{Modeling for IoDT}

In formalizing the D2XChain service registration and Drone Operator verification for IoDT, several assumptions were made:

\begin{itemize}
    \item Both the Drone Operator (DO) and Service Validator (SV) are aware of each other's public keys before service request initiation.
    \item A mock service request, akin to a CSR in traditional PKI systems, is generated within Verifpal to represent an actual service request.
    \item Token exchange between DO and SV is simplified to just signing and verification steps, assuming successful placement on the website or DNS records.
    \item The model focuses on the cryptographic interactions between a single SV and a DO, abstracting from the blockchain's consensus mechanism among multiple SVs.
\end{itemize}

The DO initiates the process with a signed service request, assuming successful token placement on the drone's infrastructure. The focus is on cryptographic operations rather than the actual mechanics of the blockchain system.

\begin{verbatim}
principal Drone_Owner[
    generates csr
    generates dronename
    certreq = CONCAT(csr, dronename)
    encreq = PKE_ENC(gb, certreq)
    signature = SIGN(a, encreq)
]
\end{verbatim}

The process of handling requests by Service Validators in D2XChain is as follows:

\begin{enumerate}
    \item The Service Validator receives an encrypted request and signature from the Drone Operator.
    \item The Validator verifies the signature and decrypts the request.
    \item The Validator creates a token ('tokentx') and appends its public key to it after a successful verification.
    \item The encrypted token, along with the Validator's signature, is then sent back to the Drone Operator.
\end{enumerate}

This process ensures secure communication and authentication between the Service Validator and the Drone Operator in the D2XChain framework.

\begin{verbatim}
principal Service_Validator[
    vrf = SIGNVERIF(ga, encreq, signature)?
    d = PKE_DEC(b, encreq)
    generates tokentx
    token = CONCAT(gb, tokentx)
    enctoken = PKE_ENC(ga, token)
    tokensig = SIGN(b, enctoken)
]
\end{verbatim}

The process of token verification and certificate publication in D2XChain is outlined as follows:

\begin{enumerate}
    \item The Service Validator receives a signed token from the Drone Operator.
    \item The Validator compares the sent token with the received one to verify their identity.
    \item Upon successful verification, the Validator signs the certificate.
    \item The signed certificate is then published to the blockchain, completing the process.
\end{enumerate}

This sequence ensures the authenticity of the Drone Operator and facilitates the secure and validated publication of certificates in the D2XChain network.

\begin{verbatim}
principal Service_Validator[
    verify = SIGNVERIF(ga, enctokendo, 
                sigtokendo)
    initialtoken = PKE_DEC(b, enctokendo)
    _ = ASSERT(initialtoken, enctoken)
]
\end{verbatim}

\subsubsection{Analysis and verification}

The security analysis using Verifpal for token exchange in D2XChain is summarized as follows:

\begin{enumerate}
    \item Verifpal is used to verify the confidentiality and authenticity of the token shared between Service Validators and Drone Operators.
    \item After signing the token with their CSR private key, Service Validators sign it with their private key, and Drone Operators host it on their IoDT platform.
    \item Service Validators then confirm the match of the received token with the initially shared token.
    \item The Verifpal analysis assumes an active attacker scenario, where the attacker can listen to and modify communications between parties.
    \item The analysis is conducted in an environment considering this active attacker, thereby verifying the robustness of the token exchange mechanism under potential security threats.
\end{enumerate}

This analysis ensures that the token exchange process in D2XChain maintains its integrity and confidentiality even in the presence of potential security attacks.

\subsection{Adversarial Model and Threat Analysis}\label{sec_8.3}

The primary objective of our framework is to establish a blockchain-based PKI system for the IoDT that is resilient against adversarial threats. In the IoDT context, our adversarial model includes entities such as Drone Operators and Service Validators. These entities may act maliciously, leading to the following potential attacks within our D2XChain framework:

\begin{itemize}
    \item IoDT Service Request (ISR) Spoofing/MITM Attack
    \item Malicious Service Validator
    \item Targeted Attack on IoDT Services
    \item Sybil Attack in IoDT Network
\end{itemize}

Table \ref{tab:table5} outlines the nature of these attacks, the category of adversaries they fall under, and how D2XChain mitigates such threats.

\begin{itemize}
    \item \textbf{ISR Spoofing/MITM Attack}: D2XChain counters service request spoofing and MITM attacks. Service Validators verify proof of ownership from the IoDT service registry, with the network's distributed nature reducing the risk of successful DNS poisoning affecting all Validators.

    \item \textbf{Malicious Service Validator}: To address dishonest actions by Service Validators, D2XChain employs smart contracts enforcing due diligence. The consensus mechanism ensures only certificates validated by a majority are accepted, and persistent dishonesty leads to removal from the network. Further research will focus on distinguishing between legitimate and malicious validators.

    \item \textbf{Victim Targeted Attack}: In cases where Service Validators neglect ISR requests for specific services or operators, D2XChain escalates these requests and ensures other Validators process them. A log of inactivity and non-participation helps identify and potentially penalize negligent Validators.

\begin{table*}[!t]
\caption{Security analysis of D2XChain against adversarial model.
\label{tab:table5}}
\centering
\begin{tabular}{llp{2in}p{2in}}\hline
Adversary Category & Attacks & Description & Mitigation\\\hline
Drone Operator & ISR Spoofing/MITM & Pretending to be a legitimate IoDT service to hijack communications. & Rigorous validation checks for ISR initiation and revocation.\\
Service Validator & Malicious Service Validator & Validators who endorse unauthorized transactions. & Consensus requirement of over 50\% of Service Validators. Monitoring of Validator activities.\\
Service Validator & Victim Targeted Attack & Overlooking ISR requests from a specific IoDT service. & Escalating transaction priority over time. Logging Validator inactivity.\\
Service Validator & Sybil Attack in IoDT & Introducing numerous deceitful Service Validators to dominate the network. & A permissioned network of trusted Service Validators. Decentralized consensus.\\\hline
\end{tabular}
\end{table*}

\item \textbf{Sybil Attack}: D2XChain employs specific strategies to mitigate the risk of Sybil attacks within the IoDT network:

\begin{itemize}
    \item \textbf{Permissioned Network}:  D2XChain operates as a permissioned network with authenticated Service Validators, each granted a certificate by D2XChain to participate actively.
    \item \textbf{Decentralized Trust}: Trust in D2XChain is established through decentralized consensus among Validators, supported by an autonomous validation protocol, rather than reliance on a single entity.
    \item \textbf{Resilience Against 51\% Collusion}: D2XChain's private and trusted Validator ensemble makes it robust against 51\% collusion attacks, a common vulnerability in blockchain systems.
    \item \textbf{Incentivization for Integrity}: The network incentivizes honest behavior among Service Validators, promoting overall integrity and reducing the likelihood of fraudulent activities.
\end{itemize}

\end{itemize}

\subsection{Assessment on Design Goals}\label{sec_8.4}

An evaluation of D2XChain against the design goals established in Section \ref{sec_5.1} yields the following:

\begin{itemize}
    \item \textbf{Compatibility with Existing Standards}: D2XChain aligns with the operational principles of traditional PKI systems, ensuring compatibility with existing IoDT infrastructures and standards.
    \item \textbf{Decentralized Trust}: The framework decentralizes trust by distributing it among multiple Service Validators rather than relying on a single Trusted Third Party, crucial for IoDT's distributed nature.
    \item \textbf{Mitigation of MITM Attacks}: D2XChain's validation protocols robustly address MITM threats, including comprehensive checks against DNS records to prevent unauthorized access.
    \item \textbf{Automation}: The system automates certificate lifecycle management, processing Service Requests as blockchain transactions with minimal human intervention.
    \item \textbf{Transparency}: Leveraging blockchain technology, D2XChain offers transparency in the auditing of IoDT service certificates, with an immutable ledger enhancing trust in the system.
\end{itemize}

\subsection{Implementation and Performance Evaluation}\label{sec_8.5}

D2XChain's implementation and performance within the IoDT are assessed as follows:

\subsubsection{Setup}\label{sec_8.5.1}

The D2XChain framework is implemented in an Ethereum blockchain environment with three key components:

\begin{enumerate}
    \item \textbf{Front-end Interface}: Developed using Vue.js, this interface is the primary interaction layer for end-users, Drone Operators, and Service Validators.
    \item \textbf{API Layer}: Built on Node.js, it connects the front-end to the blockchain, utilizing Web3js for blockchain interactions.
    \item \textbf{Blockchain Network}: Comprises Service Validator nodes, each maintaining an independent ledger copy. Smart contracts execute PoSv checks to validate transactions.
\end{enumerate}

Performance tests were performed on a typical computing setup featuring an Intel Core-i5 9600K processor, 32 GB RAM, a 500 GB SSD, and operating on Ubuntu 20.04 LTS, utilizing Python scripts for execution.

\subsubsection{Performance Analysis}\label{sec_8.5.2}

\begin{figure}
    \centering
    \includegraphics[width=\columnwidth]{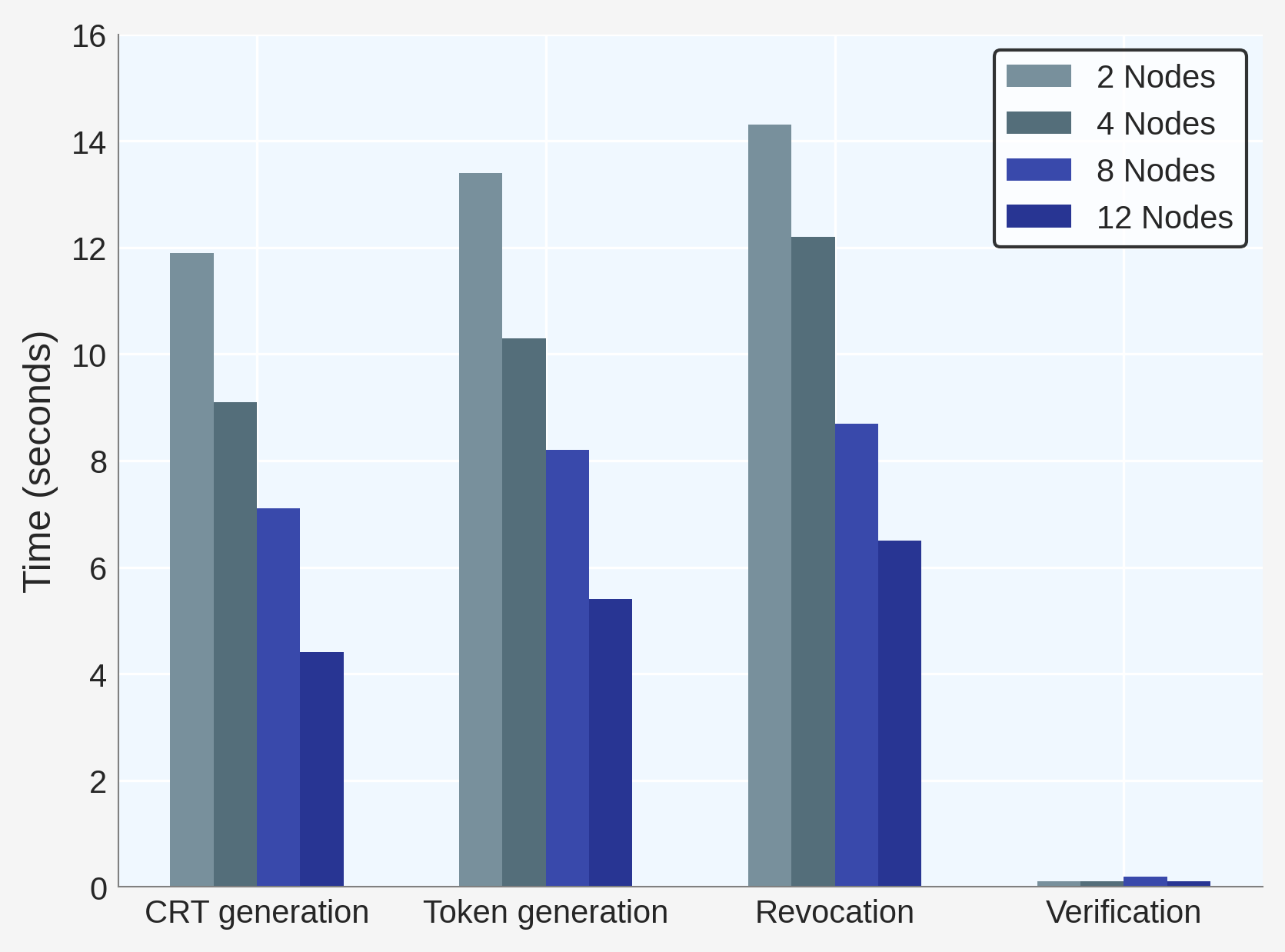}
    \caption{The latency of D2XChain operations while 2000 transactions are handled by an increasing number of SV nodes.
}
    \label{fig_9}
\end{figure}

\begin{figure}
    \centering
    \includegraphics[width=\columnwidth]{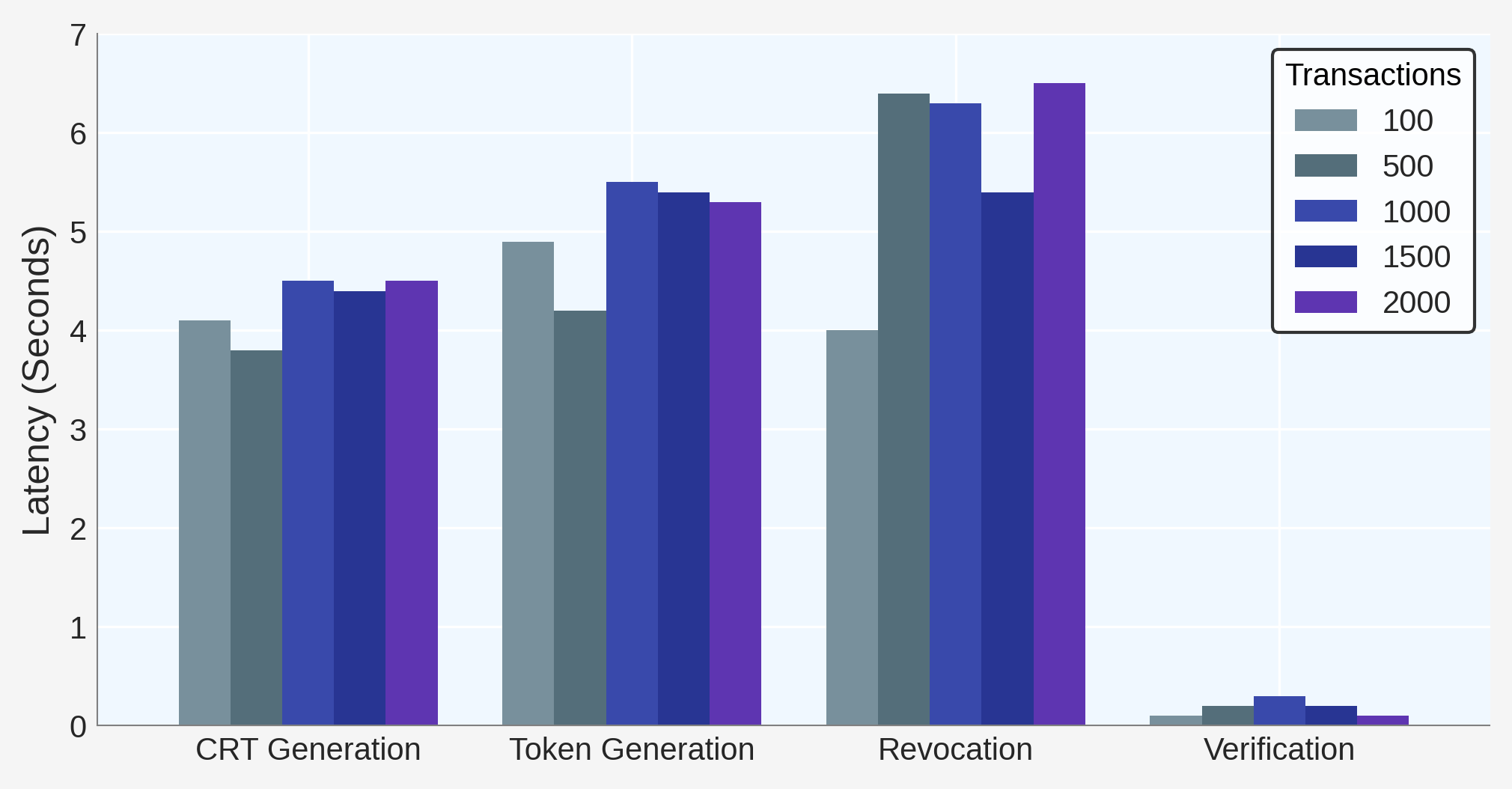}
    \caption{Latency of 12 SV nodes' D2XChain operations when there is a rise in transaction volume.
}
    \label{fig_10}
\end{figure}

\begin{figure}
    \centering
    \includegraphics[width=\columnwidth]{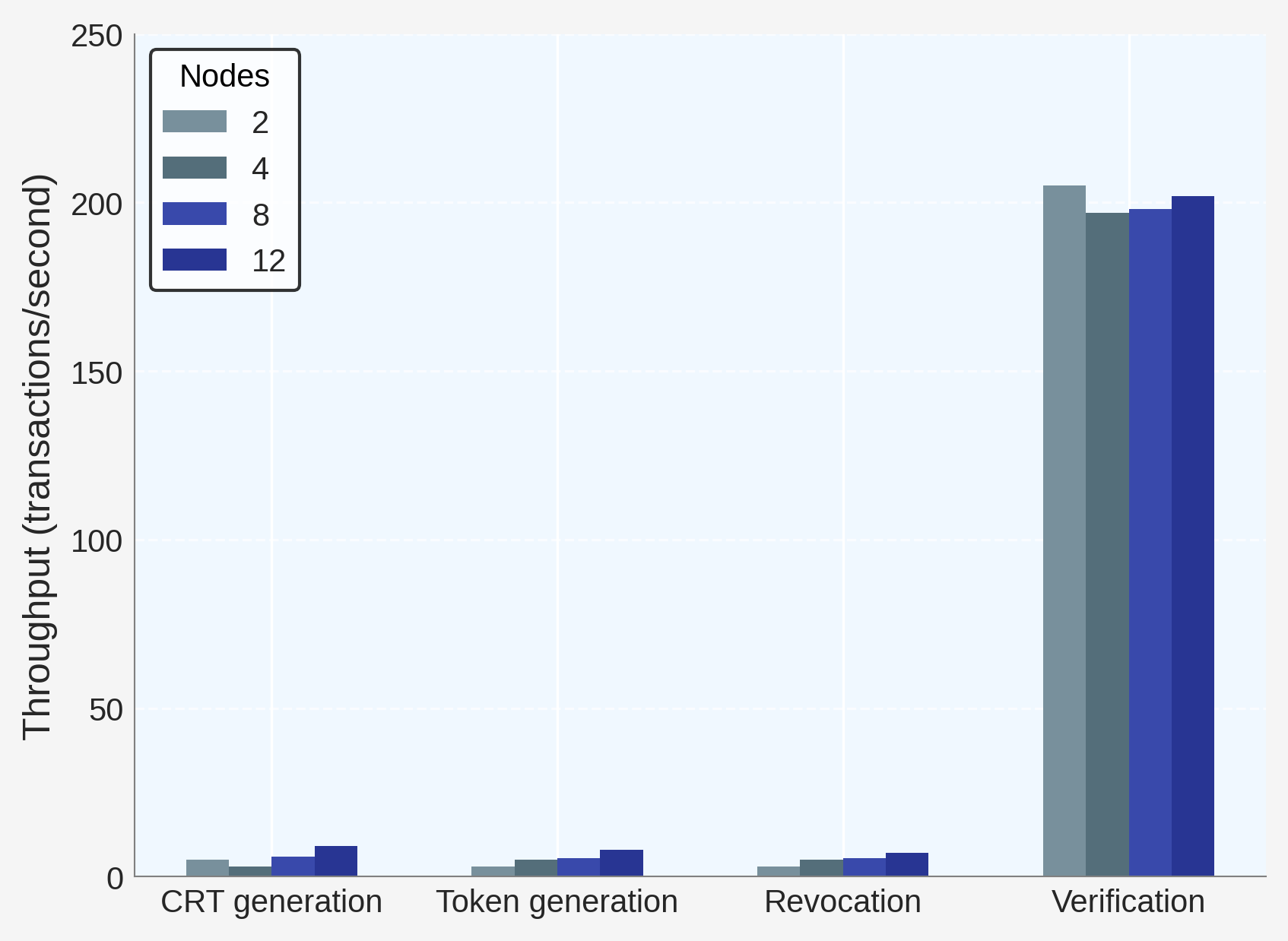}
    \caption{Throughput of D2XChain operations with 2000 transactions being handled by a growing number of SV nodes.
}
    \label{fig_11}
\end{figure}

\begin{figure}
    \centering
    \includegraphics[width=\columnwidth]{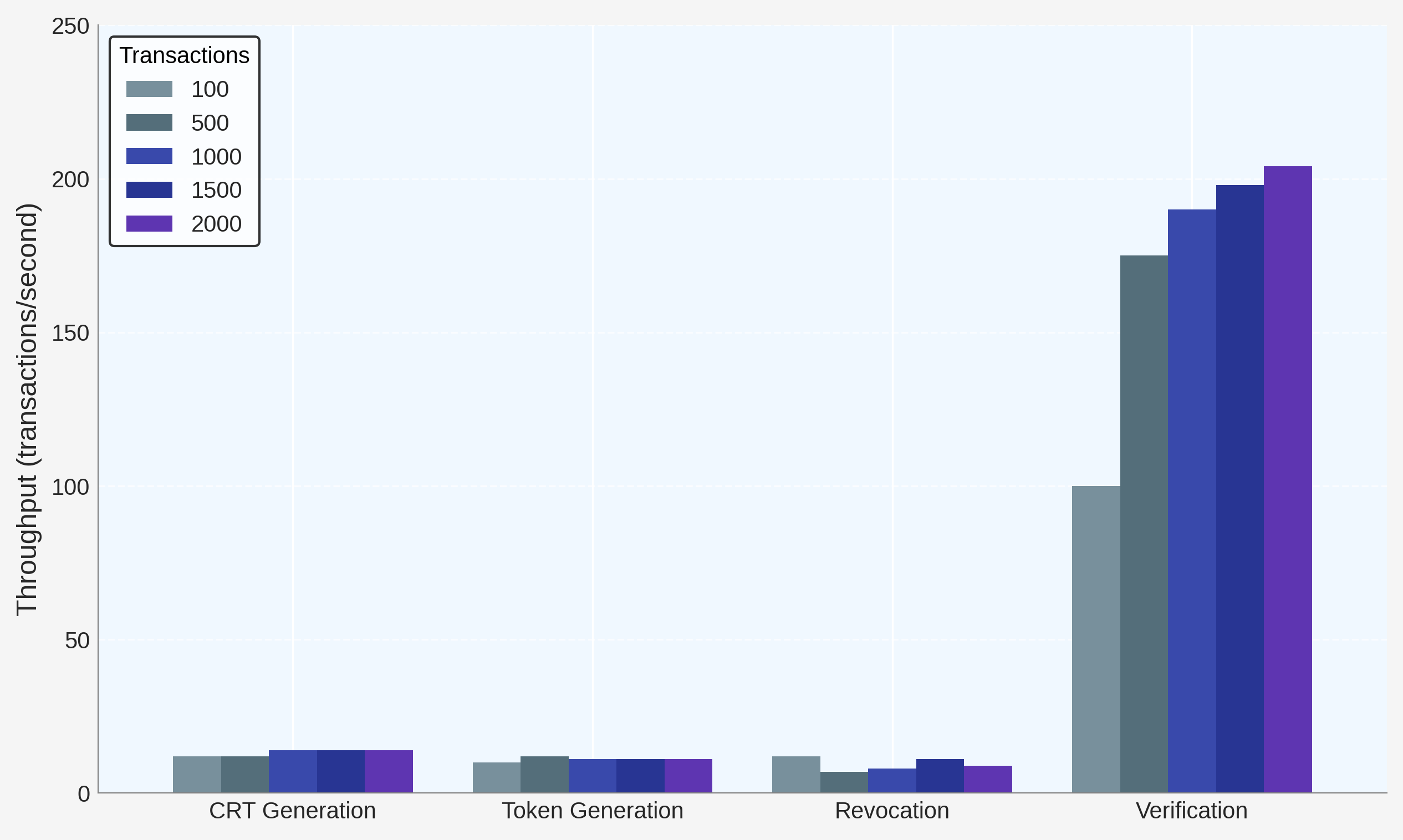}
    \caption{Throughput of 12 SV nodes' D2XChain operations with a growing amount of transactions.
}
    \label{fig_12}
\end{figure}

The performance of D2XChain was evaluated under varying conditions:

\begin{itemize}
    \item \textbf{Experimental Setup}: Two scenarios were tested—varying the number of nodes with a fixed number of transactions, and vice versa. The registration mechanism (token issuance and CDT generation), revocation, and verification processes were analyzed separately.
    
    \item \textbf{Results with Varying Nodes}: Latency and throughput were assessed with up to 12 nodes and a fixed transaction count of 2000. The results (Figures \ref{fig_9} and \ref{fig_11}) showed a decrease in latency and an increase in throughput for registration and revocation with more nodes. Verification throughput remained relatively constant.
    
    \item \textbf{Results with Varying Transactions}: The second setup, with a constant node count and escalating transactions, indicated a throughput decline in registration and revocation but an increase in verification operations (Figures \ref{fig_10} and \ref{fig_12}). There were differences in the patterns of latency: higher for revocation and registration, lower for verification for more than 500 transactions.

\end{itemize}

The investigation highlights D2XChain's usefulness in an IoDT context by showing that it can strike a balance between throughput and the number of SV nodes.

D2XChain's approach to certificate issuance in the IoDT context is evaluated considering the expected rise in demand:

\begin{itemize}
    \item \textbf{Decentralized Operation}: D2XChain enables multiple Service Validators (SVs) to process certificate issuance requests concurrently, enhancing throughput.
    
    \item \textbf{Performance Metrics}: In tests with 12 SV nodes handling 2000 transactions, the throughput for certificate registration and validation was around 10 seconds. This indicates effective handling of increased certificate issuance frequencies.
    
    \item \textbf{Scalability}: The number of SV nodes in D2XChain can be scaled to improve performance, countering potential limitations related to blockchain size or transaction rates.
    
    \item \textbf{Adaptability to Ethereum Advancements}: Ethereum features like sharding, roll-ups, state channels, and the PoS protocol can help deal with the difficulties associated with often issuing certificates for IoDT devices. D2XChain can integrate with these features without extra costs or complexity.
    
    \item \textbf{Certificate Validity Status}: The framework aggregates certificate validity status into a vector, streamlining the verification process by eliminating the need to traverse the entire certificate chain.
\end{itemize}

\begin{table*}[!t]
    \caption{D2XChain's cost comparison for IoDT services vs the conventional CA approach.
\label{tab:table6}}
    \centering
    \begin{tabular}{lp{0.5in}p{0.5in}p{0.75in}p{0.75in}p{1.5in}}\hline
    Operation & Gas Used (Wei) & Gas Price (GWei) & Total Cost (Eth) & Total Cost (USD) & Traditional IoDT Certificate Price (USD)\\\hline
    CRT Generation & 61603 & 1 & 0.0000616 & \$0.04 & \\
    Certificate Registration & 64579 & 1 & 0.0000646 & \$0.05 & approx \$10 per year\\
    Certificate Revocation & 35341 & 1 & 0.0000353 & \$0.03 & 0\\
    Certificate Validation & 0 & 1 & 0.0000000 & \$0.00 & 0\\\hline
    \end{tabular}
\end{table*}

\subsubsection{Cost analysis}

\begin{table*}[htbp]
    \caption{Assessment of IoDT-Related PKI Techniques Compared to D2XChain\label{tab:revisedTable7}}
    \begin{center}
    \begin{tabular}{lcccc}
    \hline
    Evaluation Metric & D2XChain & CertLedger & BlockPKI & CloudX509 \\
    \hline
    Cost for Issuing Certificates (in Ether) & 0.0001262 & - & 0.07 & - \\
    Client Storage Requirements (MB/annum) & 10 & 128 & - & - \\
    Average Block Size (in KB) & 0.564 & 128 & - & 1 \\
    Overhead for TLS Handshake (Bytes per 1000 certs) & 0 & 5657 & - & - \\
    Time to Issue Certificates (Sec per 1000 certs) & 9 & - & 120 & 2040 \\
    Blockchain Size on Annual Basis (in GB) & 86.67 & 512 & - & - \\
    Availability of a Working Model & Yes & Yes & No & No \\
    \hline
    \end{tabular}
    \end{center}
\end{table*}

D2XChain's deployment in the IoDT ecosystem on a private Ethereum network reveals its cost-effectiveness and efficiency:

\begin{itemize}
    \item \textbf{Operational Costs}: D2XChain incurs modest operational costs of \$0.09 for certificate registration and \$0.03 for revocation, significantly lower than traditional PKI systems. This cost advantage, coupled with mitigation of the single point of failure issue, makes D2XChain an economically viable alternative.
    
    \item \textbf{Reward Mechanism for SVs}: An incentive scheme is put up for Service Validators, whereby certificate creation costs are distributed as compensation for carrying out sanity tests. This study does not address the intricacies of the block rewards distribution.
    
    \item \textbf{Blockchain Configuration and Costs}: Both private and public blockchains can use the framework; public transactions will cost money, while private networks will need an initial investment but may not charge transaction fees.

    \item \textbf{Storage Analysis}: A theoretical model of the D2XChain ledger was analyzed for storage efficiency. The analysis, using methodologies from CertLedger, shows D2XChain's superior performance in storage efficiency and processing time compared to similar frameworks, as detailed in Table \ref{tab:revisedTable7}.
\end{itemize}

\begin{table*}[!t]
    \caption{D2XChain smart contracts security vulnerabilities detection in the IoDT context.\label{tab:table8}}
    \centering
    \begin{tabular}{|c|c |c| c|}\hline
    Vulnerability class & Vulnerability group & Vulnerability & Detection in D2XChain smart contracts\\\hline

    & & Vulnerable off-chain server & \cmark\\
    & \multirow{-2}{*}{Trust} & Overpowered owner & \xmark\\\cline{2-4}
    
    & Privacy & Sensitive data visibility & \xmark\\\cline{2-4}

    & & Constructor name & \xmark\\
    & & Authorization with tx.origin & \xmark\\
    & & Suicidal contracts & \xmark\\
    & \multirow{-4}{*}{Authorization} & Generous contracts & \xmark\\\cline{2-4}

    & & Game Theory issues & \xmark\\
    & & Tokenomics issues & \xmark\\
    \multirow{-10}{*}{Model} & \multirow{-3}{*}{Economy} &Voting issues & \xmark\\\hline

    & & Overlap attack & \xmark\\
    & & Delegatecall and storage layout & \cmark\\
    & \multirow{-3}{*}{Storage access} & Uninitialized storage pointer & \xmark\\\cline{2-4}
    
        & & Over/underflow & \xmark\\
    & \multirow{-2}{*}{Arithmetic} & Precision issues & \xmark\\\cline{2-4}

    & & Assembly return in constructor & \xmark\\
    & & Function type variable jump & \xmark\\
    \multirow{-8}{*}{Language}& \multirow{-3}{*}{Internal control flow} & Multiple inheritance & \xmark\\\hline

        & & Transaction censorship & \xmark\\
    & & Random with blockhash & \xmark\\        
    & & Timestamp manipulation & \xmark\\
    & \multirow{-4}{*}{Block content manipulation} & Front-running/transaction reordering & \xmark\\\cline{2-4}
    
        & & Signature collisions & \xmark\\
    & \multirow{-2}{*}{Message structure} & Short address attack & \xmark\\\cline{2-4}

        & & Infinite loops & \xmark\\
    & \multirow{-2}{*}{Contract interaction} & Transfer provides too little gas & \xmark\\\cline{2-4}

        & & DoS with selfdestruct & \xmark\\
    & & DoS with revert & \xmark\\        
    & & Reentrancy & \xmark\\
    & \multirow{-4}{*}{Contract interaction} & Unchecked low-level call & \xmark\\\cline{2-4}

    & & Present Ether & \xmark\\
    & & Ether transfer with mining & \xmark\\
    \multirow{-15}{*}{Blockchain}& \multirow{-3}{*}{Ether transfer} & Ether transfer with selfdestruct & \xmark\\\hline

    \end{tabular}
    \end{table*}

\subsubsection{Smart Contract Vulnerabilities in IoDT}

The security of smart contracts in D2XChain is crucial for maintaining the integrity of the blockchain-based system:

\begin{itemize}
    \item \textbf{Importance of Smart Contract Integrity}: Given their immutable nature, ensuring the security of smart contracts is essential in the IoDT framework.
    
    \item \textbf{Analytical Tools for Vulnerability Detection}: Various tools can analyze smart contracts, examining either EVM bytecode or source code. D2XChain employed SmartCheck, an open-source tool that analyzes Solidity source code for vulnerabilities.
    
    \item \textbf{SmartCheck Analysis}: SmartCheck categorizes vulnerabilities into several classes and applies numerous rules for detection. D2XChain's smart contracts, totaling 404 lines of code, were analyzed against approximately 80 rules, as summarized in Table \ref{tab:table8}.
    
    \item \textbf{Identified Vulnerabilities}: Vulnerabilities related to off-chain server connections were found, due to interfacing with an oracle to fetch tokens from DNS records. Issues with Delegatecall and Storage layout were also noted, attributed to the loading of data from memory using assembly code.
    
    \item \textbf{System Robustness}: The analysis indicates that D2XChain is robust against a wide range of security issues, ensuring the integrity of decentralized trust management in the IoDT ecosystem.
\end{itemize}

\section{Conclusion}\label{sec_9}

D2XChain is a pioneering blockchain-based PKI framework designed specifically for the IoDT, utilizing a private Ethereum blockchain with a PoSv protocol. It achieves decentralized trust management while complying with the X.509 standard, supporting essential PKI operations like registration, validation, revocation, and verification. Enhanced by smart contracts with thorough sanity checks, D2XChain effectively fortifies against adversarial threats, ensuring robust security in the IoDT environment. Performance evaluations affirm its aptitude for real-time IoDT applications, marking a substantial advancement in secure, decentralized PKI for drone services. Future research aims to refine the system further, focusing on integrating PoSv into the blockchain's consensus mechanism and developing a dedicated ledger optimized for IoDT PKI operations, thereby elevating the security and functionality in the dynamic IoDT landscape.


\bibliographystyle{ieeetr}
\bibliography{refs}

\end{document}